\documentclass[a4paper,fleqn,authoryear]{cas-sc}
\pdfoutput=1 
\usepackage[square,sort,comma,numbers]{natbib}

\usepackage{multirow}
\usepackage{xcolor}

\def\tsc#1{\csdef{#1}{\textsc{\lowercase{#1}}\xspace}}
\tsc{WGM}
\tsc{QE}
\tsc{EP}
\tsc{PMS}
\tsc{BEC}
\tsc{DE}

\makeatletter

\renewcommand{\@oddfoot}{}
\renewcommand{\@evenfoot}{}
\makeatother

\usepackage{placeins}
\usepackage{pmboxdraw}
\usepackage{lscape}
\usepackage[dvipsnames]{xcolor}
\usepackage{algorithm}
\usepackage{algorithmic}
\usepackage{caption}
\usepackage{changepage}
\usepackage{tabularx}
\usepackage{array}
\usepackage{natbib}

\usepackage{lipsum}

\makeatletter
\def\ps@pprintTitle{%
 \let\@oddhead\@empty
 \let\@evenhead\@empty
 \def\@oddfoot{}%
 \let\@evenfoot\@oddfoot}
\makeatother

\usepackage{mathtools}

\begin{document}
\let\WriteBookmarks\relax
\def\floatpagepagefraction{1}
\def\textpagefraction{.001}

\shorttitle{Climate Health Nexus}

\shortauthors{ Kaur et al.}

\title [mode = title]{EpiClim: Weekly District-Wise all-India multi-epidemics Climate-Health Dataset for accelerated GeoHealth research}

\author[1]{ Gurleen Kaur} 
\ead{gurleen21@iiserb.ac.in}
\affiliation[1]{organization={Indian Institute of Science Education and Research, Bhopal, India }}
\author[2]{ Shubham Ghoshal} 
\ead{shubhamghosal100@gmail.com }
\affiliation[2]{organization={Banaras Hindu University }}

\author[3]{Reena Marbate} 
\ead{reenamarbate14796@gmail.com}
\affiliation[3]{organization={Indian Institute of Public Health, Delhi, India }}

\author[4]{Neetiraj Malviya} 
\ead{neetirajmalviya@gmail.com}
\affiliation[4]{organization={Centre for Development of Advanced Computing, Pune, India }}
\author[5]{Arshmehar Kaur} 
\ead{arshmeharkaur05@gmail.com}
\affiliation[5]{organization={Guru Tegh Bahadur Institute of Technology, Delhi, India }}
\author[6]{Vaisakh SB} 
\ead{ vaisakh.sb@tropmet.res.in}
\affiliation[6]{organization={Indian Institute of Tropical Meteorology, Pune, India }}

\author[7]{Amit Kumar Srivastava} 
\ead{AmitKumar.Srivastava@zalf.de }
\affiliation[7]{organization={Leibniz-Centre for Agricultural Landscape Research (ZALF), Müncheberg, Germany}}

\author[8]{Manmeet Singh} 
\ead{manmeet.singh@utexas.edu }
\affiliation[8]{organization={ Jackson School of Geosciences, University of Texas at Austin, Austin, TX, USA}}

\cortext[cor2]{Correspondence to: \textcolor{black}{Indian Institute of Science Education and Research, Bhopal, India}}

\begin{abstract} 
Climate change directly affects public health as it significantly influences the emergence and spread of epidemics. Factors such as temperature, precipitation majorly impact and promote the growth of prevalent diseases transmitted by vectors and water-borne diseases as seen from historical and contemporary evidence. We will see in this research work the correlation of various diseases like dengue, cholera , acute diarrheal disease and malaria with temperature for example dengue and malaria have a good correlation with temperature and cholera on the other hand usually outbreaks with anomalies with precipitation showing a good correlation with the precipitation factor. Collection of high quality data has become a must to be able to foster the growth of climate health models in understanding and predicting disease outbreaks due to climate variability.  Innovative developments in weather and climate science, especially through AI-enabled numerical weather prediction, AI-NWP, boost confidence in forecasting key variables such as temperature and precipitation. However, a major obstacle to the integration of climate models with health prediction systems lies in the inaccessibility of comprehensive, publicly accessible health datasets for various diseases. Such data sets, especially on granular spatial and temporal scales, are crucial to advance climate health research and model development. In this paper we present the first weekly district-wise dataset for major epidemics spanning across India from 2009 to 2022. The data utilized in this study has been sourced from the Integrated Disease Surveillance Programme(IDSP),a publically available database maintained by the Government of India. This dataset, named "EpiClim: India's Epidemic-Climate Dataset", bridges the gap by providing detailed insights into health data tailored for climate-health modeling. The data set offers insight into the temporal and spatial dynamics of diseases such as dengue, malaria, and acute-diarrheal disease.  Such datasets open the gates to integrate climate forecasts with epidemic prediction models and create actionable insights for policymakers, public health officials, and researchers. This work lays the foundation for coupling predictive climate health models with numerical weather and climate models, driving an innovation that seeks to understand and mitigate public health crises caused by climate.

Dataset link : \href{https://zenodo.org/records/14580510}{DOI for EpiClim}

\end{abstract}

\maketitle

\section{Introduction}
Climate change and its related threats are growing at an alarming rate. By disrupting natural ecosystems and weather patterns climate change poses a serious threat to human health, eventually leading to climate-induced epidemics. It causes extreme weather conditions such as droughts, record breaking heat waves on land and in the ocean, drenching rains, severe floods, extreme wildfires, and widespread flooding during hurricanes which directly or indirectly affects human health. These extreme weather conditions favor the growth of various kinds of epidemic diseases. According to the World Health Organization (WHO, 2024), Climate change is projected to significantly impact global health, leading to an estimated 250,000 additional deaths annually between 2030 and 2050 due to malnutrition, malaria, diarrhea, and heat-related illnesses. These climate-sensitive diseases are expected to worsen as rising temperatures and extreme weather conditions create more favorable environments for their spread. Studies have found that long-term climate warming tends to favor the geographic expansion of several infectious diseases (Epstein et al., 1998, Ostfeld and Brunner, 2015, Rodó et al., 2013), and that extreme weather events may help create opportunities for more clustered disease outbreaks or outbreaks at non-traditional places and time (Epstein, 2000). Warmer climates expand the geographic range where vectors like mosquitoes can survive and breed, while increased rainfall creates more standing water, providing additional breeding grounds. Even droughts can contribute by forming stagnant water pools from previously flowing sources. Additionally, climate change extends disease transmission seasons by creating more favorable environmental conditions for pathogens. Temperature changes can also alter vector behavior, such as increasing mosquito biting activity, which can reduce the effectiveness of protective measures. According to IPCC(AR6) reports the occurrence of climate-related food-borne and water-borne diseases (very high confidence) and the incidence of vector-borne diseases (high confidence) have increased. In the assessed regions, some mental health challenges are associated with increased temperatures (high confidence), trauma from extreme events (very high confidence), and loss of livelihoods and culture (high confidence).

The role of climate change in altering the distribution and behavior of disease vectors is well documented. For instance, the expansion of mosquito habitats due to warmer temperatures has been linked to the resurgence of diseases like dengue and chikungunya in previously unaffected regions (Rocklöv et al., 2016). Similarly, changes in precipitation patterns have been associated with the spread of waterborne diseases such as cholera and leptospirosis, particularly in areas with inadequate sanitation infrastructure (Levy et al., 2016). These findings highlight the need for a deeper understanding of the environmental drivers of disease transmission to inform effective public health interventions.

In addition to infectious diseases, climate change is also contributing to the rise of noncommunicable diseases (NCDs). Prolonged exposure to high temperatures has been linked to cardiovascular and respiratory illnesses, while air pollution exacerbated by climate change is a leading risk factor for chronic conditions such as asthma and lung cancer (Patz et al., 2014). Vulnerable populations, including the elderly, children, and those with preexisting health conditions, are disproportionately affected by these climate-related health risks (Haines et al., 2006).

The economic burden of climate change on health systems is also significant. The costs associated with treating climate-sensitive diseases, rebuilding healthcare infrastructure damaged by extreme weather events, and addressing the long-term health impacts of climate change are substantial (WHO, 2018). These challenges are further compounded by the limited capacity of healthcare systems in low- and middle-income countries to respond to the growing health threats posed by climate change (Ebi et al., 2021). Climate change is transforming the socioeconomic and health landscapes of communities around the world(Borghi, J., \& Kühn, M. (2024)). The increasing frequency and intensity of extreme weather events, such as hurricanes, cyclones, and wildfires, have been linked to significant health burdens, including injuries, displacement, and mental health disorders (Watts et al., 2019). Furthermore, the disruption of agricultural systems due to erratic rainfall and prolonged droughts has led to food insecurity, malnutrition, and the exacerbation of poverty, particularly in vulnerable regions (Myers et al., 2017). These cascading effects underscore the interconnectedness of climate change, health, and socioeconomic stability(Nie, P., Zhao, K., Ma, D., Liu, H., Amin, S., \& Yasin, I. (2024)).

 Despite the growing recognition of the impacts of climate change there is a lack of collection and analysis of data to help mitigate climate induced health problems(Garcia, E., Eckel, S. P., Silva, S., McConnell, R., Johnston, J., Sanders, K. T., Habre, R., \& Baccarelli, A. (2024)). By integrating robust health datasets with climate variables, researchers can develop predictive models to assess the onset and spread of climate-sensitive diseases. Climatic variables like temperature and precipitation increase the dynamics of diseases, which intensify the effects of epidemics such as malaria, dengue, and acute diarrheal diseases(Athirsha, A., Anitha, M., Sundar, J. S., Kalpana, S., Valarmathi, S., \& Srinivas, G. (2024)). Nevertheless, advancement in the science of weather and climate, especially in artificial intelligence-enabled numerical weather prediction (AI-NWP), calls for mitigation and prediction of climate-induced epidemics.
There remains a critical gap in the availability of high-quality, granular data to support the development of predictive models and targeted interventions. The integration of climate and health data is essential for identifying vulnerable populations, forecasting disease outbreaks, and designing adaptive strategies to mitigate the health impacts of climate change (Campbell-Lendrum et al., 2015). Advances in data science, machine learning, and geospatial technologies offer promising opportunities to address these challenges, enabling the creation of robust climate-health models that can inform policy and practice (Birkmann et al., 2021). Such models require robust health data sets integrated with climate variables. However, the lack of detailed publicly available epidemiological data impedes progress in this domain(Fairchild, G., Tasseff, B., Khalsa, H., Generous, N., Daughton, A. R., Velappan, N., Priedhorsky, R., \& Deshpande, A. (2018)). To bridge this gap, we present a weekly epidemiological data set for India spanning from 2009 to 2022. This dataset captures the dynamics of multiple diseases, enabling spatial and temporal analyses and facilitating the development of climate health models. This data set aims to support the development of robust climate health prediction models to mitigate the impacts of climate change on public health. These tools are essential for protecting public health and improving quality of life in the context of climate and health challenges. Our work aims to:

\begin{enumerate}
    \item Investigate the role of climatic factors in disease spread.
    \item Presenting an analytical report for the developed dataset providing in-depth insights and conclusions for epidemic diseases
    \item Provide a comprehensive dataset for public health research.
    \item Advance the integration of climate models with health prediction systems to enable informed public health interventions.
\end{enumerate}

\section{Data Description and Technical Validation}

\subsection{Dataset Overview}
Our dataset- EpiClim was scrupulously compiled from open sources, with a significant contribution from the Integrated Disease Surveillance Programme (IDSP) portal \footnote{\url{https://idsp.mohfw.gov.in/index4.php?lang=1&level=0&linkid=406&lid=3689}}. IDSP publishes weekly reports on various disease outbreaks across India. From such reports, district-wise health data for Dengue, Malaria, Cholera, and Acute Diarrhoeal Disease was retrieved. Names of districts were linked to their respective geographic coordinates- latitude and longitude-for spatial analysis. The climate data was taken from era 5 using Google Earth engine and was mapped with the same cities and the same time period as the health data taken from IDSP portal.

We incorporate relevant climatic variables , including temperature, precipitation, and LAI from publicly available climate datasets to complement health information like number of deaths, name of the disease and number of cases in a week. It was ensured that those variables are included based on potential influences suggested in the existing literature regarding the dynamics of diseases. The combination of health and climatic data has fostered the construction of a robust dataset ready to investigate the complex interactions between climate and health.

\subsection{Data Summary}
Table~\ref{table:dataset_summary} and Table~\ref{table:dataset_summary2} summarize the main diseases and states by cases, along with the statistics of the climatic factors. The most prevalent disease recorded in the dataset is Acute Diarrhoeal Disease, followed by Dengue and Cholera. West Bengal reports the highest number of cases among the states. Climatic factors such as temperature and precipitation show significant variations, with average temperatures around 304.5K and mean daily precipitation of 0.46mm.

\newpage
\section*{Table 1: Excerpts from the Epiclim dataset for the period 2009-2022 in India}

\begin{table}[h]
    \centering
    \renewcommand{\thetable}{1(a)} 
    \begin{minipage}{0.45\textwidth}
        \centering
        \caption{Disease cases in India}
        \label{table:dataset_summary}
        \begin{tabular}{|l|r|}
            \toprule
            \textbf{Disease} & \textbf{Cases} \\
            \midrule
            Acute Diarrheal Disease & 251,456 \\
            Dengue & 238,047 \\
            Cholera & 126,495 \\
            Malaria & 111,858 \\
            Chikungunya & 53,289 \\
            \bottomrule
        \end{tabular}
    \end{minipage}%
    \hfill
    \addtocounter{table}{-1} 
    \renewcommand{\thetable}{1(b)} 
    \begin{minipage}{0.45\textwidth}
        \centering
        \caption{Top States with the highest number of disease cases}
        \label{table:dataset_summary2}
        \begin{tabular}{|l|r|}
            \toprule
            \textbf{State/UT} & \textbf{Cases} \\
            \midrule
            West Bengal & 178,830 \\
            Delhi & 80,933 \\
            Uttar Pradesh & 60,173 \\
            Maharashtra & 53,575 \\
            Karnataka & 39,532 \\
            \bottomrule
        \end{tabular}
    \end{minipage}
\end{table}

\section{Methodology}

The approach we undertook for our research includes the collection of data, curation of the dataset, and visualization of our findings catering to disease outbreak data integrated with climatic variables. 

The epidemiological data was gathered by web scraping and sourced mainly from the \textbf{Integrated Disease Surveillance Programme (IDSP)}, which is a nationwide disease surveillance system in India. This data was combined with climate variables, and the climate data was primarily obtained from \textbf{Google Earth Engine}. The epidemiological data was cross-referenced with climate datasets to ensure accuracy and consistency.

\subsection*{Dataset Description}
The dataset has 8984 rows and includes 15 features covering epidemiological, geographic, and climatic data. Notable variables include:

\begin{itemize}
    \item \textbf{Temporal Data:} Week, month, and year of outbreaks.
    \item \textbf{Geographic Data:} State, district, latitude, and longitude.
    \item \textbf{Epidemiological Data:} Number of diseases, cases, and deaths.
    \item \textbf{Climatic Data:} Precipitation (preci), Leaf Area Index (LAI), and temperature (Temp).
\end{itemize}

\subsection*{Data Integration and Framework}
We developed a comprehensive framework to analyze the relationship between precipitation, temperature, and vegetation health at different latitudes and longitudes using \textbf{Google Earth Engine} and \textbf{IDSP} data, iterating the two to create a novel dataset. Data was collected, and different variables were taken into account, such as LAI, temperature, precipitation, number of diseases, cases, and deaths from the \textbf{Integrated Disease Surveillance Programme (IDSP)}, which is a nationwide disease surveillance system in India.

The spatial data was collected from the \textbf{ECMWF ERA5-Land dataset} using \textbf{Google Earth Engine}, which provides high-resolution hourly climate variables, including total precipitation. The latitude and longitude were extracted for various districts and states in India over a defined period (2009--2022) at a weekly resolution. The precipitation data was mapped with epidemiology data based on latitude and longitude and combined with variables such as LAI, temperature, precipitation, number of diseases, cases and deaths.
The data pre-processing phase, focused on conversion to a format apt for analysis-in particular, where string values might need to be numerically formatted if possible like from "Cases" to "Integer" and missing values have to be processed e.g., the NaN value within the column of "Deaths"-the ultimate was a big amount of visualisation to squeeze actionables. Spatial distribution maps for 2022 were created using Python to spot hotspots for diseases, and temporal trend analyses were carried out to study Acute Diarrheal Disease across the years. Year-wise subplots have been used to comparatively plot selected diseases throughout the years like 2011, 2013, and 2015. This will help in finding spatial and temporal trends, which will help in giving an idea to build predictive climate-health models. Further details on technical validation and coding will detail this methodology.

\subsection*{Geospatial Mapping}
\begin{figure*}[h]
    \centering
    \includegraphics[width=0.8\textwidth]{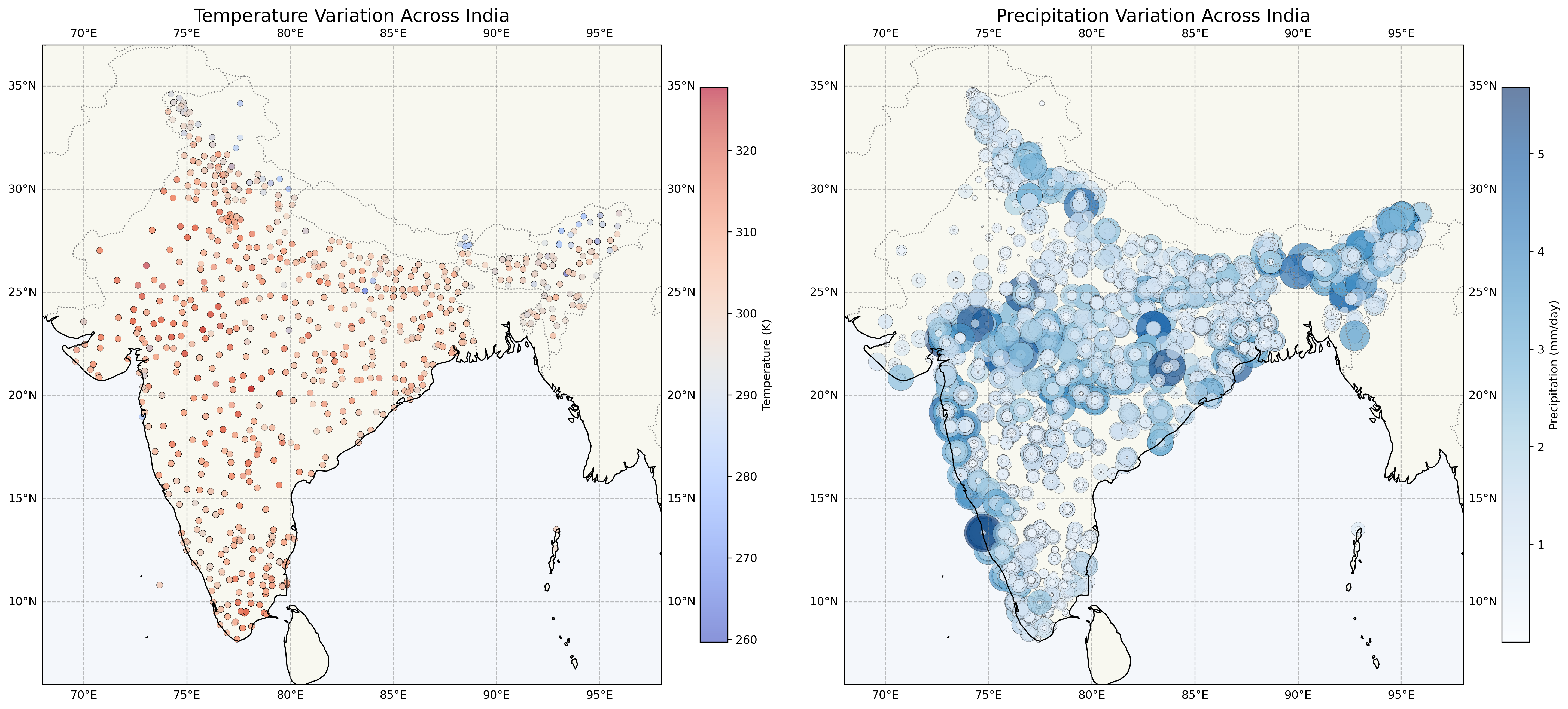}
    \caption{Geo-spatial variations in temperature (left) and precipitation (right) across India.}
    \label{fig:climate_maps}
\end{figure*}

The geospatial mapping of outbreak locations enhances the understanding of the spread patterns and helps design targeted response strategies by analyzing disease spread trends. By gathering and integrating these variables into datasets, we ensured a structured approach to understanding how climatic and environmental conditions influence disease prevalence. Figure~\ref{fig:climate_maps} illustrates the geospatial mapping of temperature and precipitation variations across India, providing insights into climate-related patterns.

\subsection*{Data Visualization}
\begin{figure}[h]
    \centering
    \includegraphics[width=0.5\columnwidth]{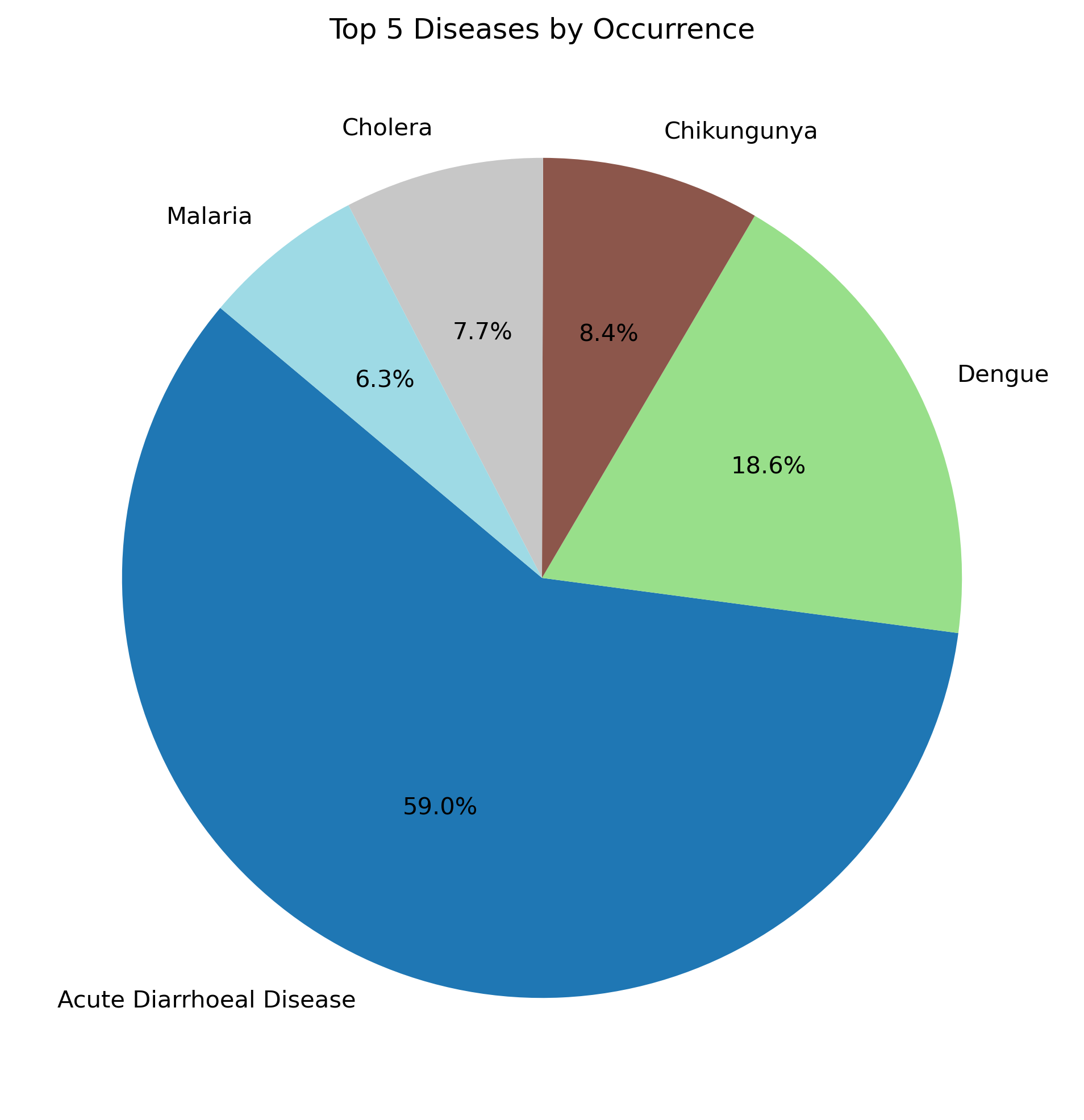}
    \caption{Distribution of top 5 diseases by reported cases for the year 2009 to 2022 across India. Acute Diarrheal Disease accounts for the largest proportion.}
    \label{fig:disease_pie}
\end{figure}

After data collection, we performed various visualizations of the data to understand it on a deeper level.
The methodology for this project involved a systematic approach to collecting, processing, and visualizing epidemiological data, integrated with relevant climatic variables. First, the epidemiological data was gathered from publicly available reports and cross-referenced with climate datasets to ensure accuracy and consistency.  Figure~\ref{fig:disease_pie} illustrates the distribution of top 5 diseases reported from 2009 to 2022 accross India.
\clearpage

\subsection{Visualization and Analysis}

The dataset is enriched with key visualizations that highlight its scope and applicability. The codebase for the visualisation can be found at \href{https://github.com/Gurleen110011/Climate_Health_Nexus}{GitHub Repository} :
\begin{enumerate}
    \item \textbf{Temporal Trends of Cases and Deaths: }  
    Figure ~\ref{fig:deaths_timeseries} illustrates the temporal dynamics of reported cases and deaths, smoothed over a 100-day moving average. Both metrics demonstrate periodic peaks, indicative of seasonality in epidemic occurrences.

    \begin{figure*}[h]
    \centering
    \includegraphics[width=0.8\textwidth]{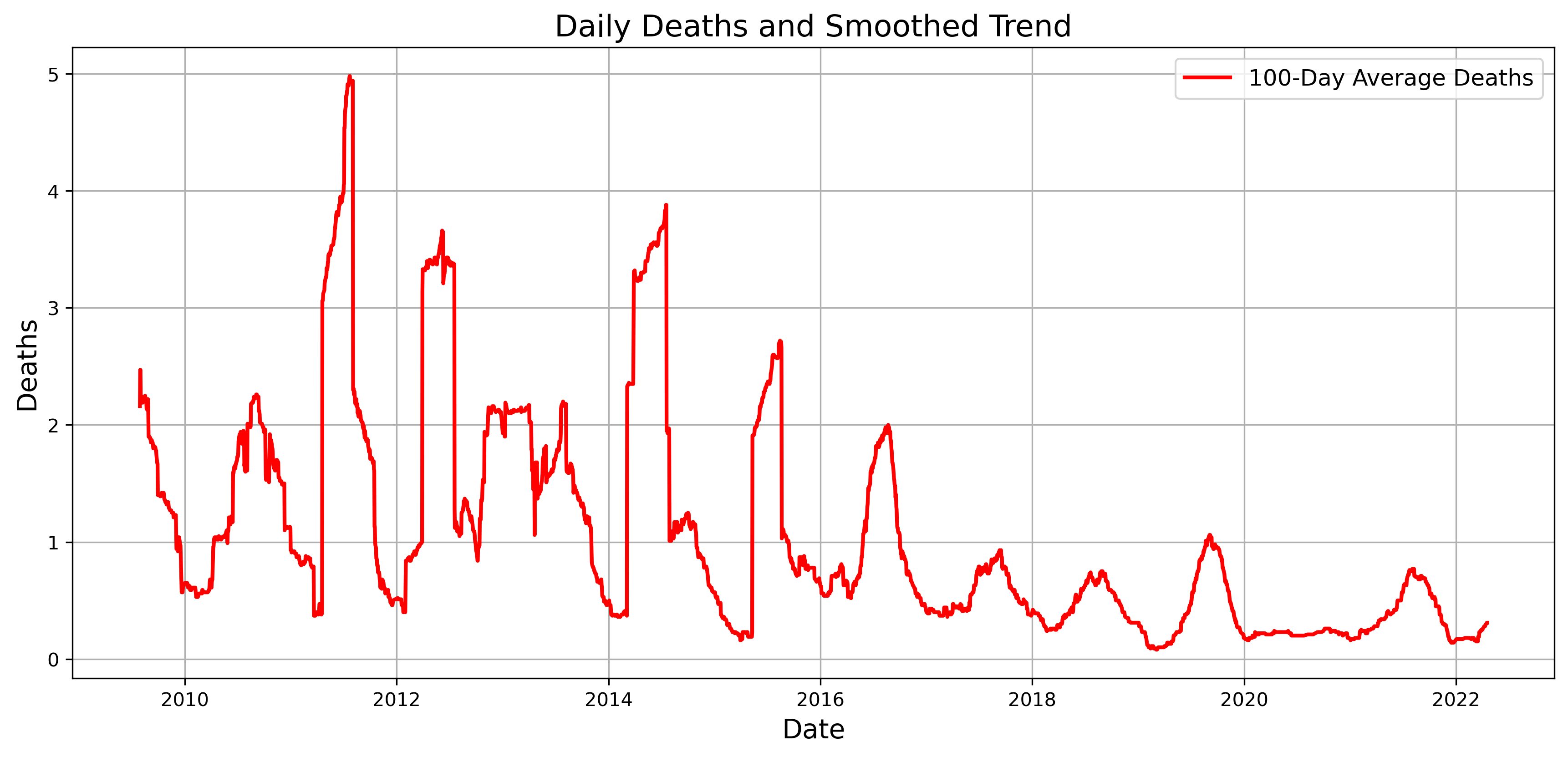}
    \caption{Temporal trends of daily reported deaths, smoothed over a 100-day average.}
    \label{fig:deaths_timeseries}
    \end{figure*}

    \item \textbf{Disease Prevalence Distribution:} The proportional distribution of top diseases is depicted in Figure~\ref{fig:disease_pie}. Acute diarrheal disease dominates the data set, comprising 59\% of the total cases.
\end{enumerate}

\begin{figure}[h]
    \centering
    \includegraphics[width=0.5\textwidth]{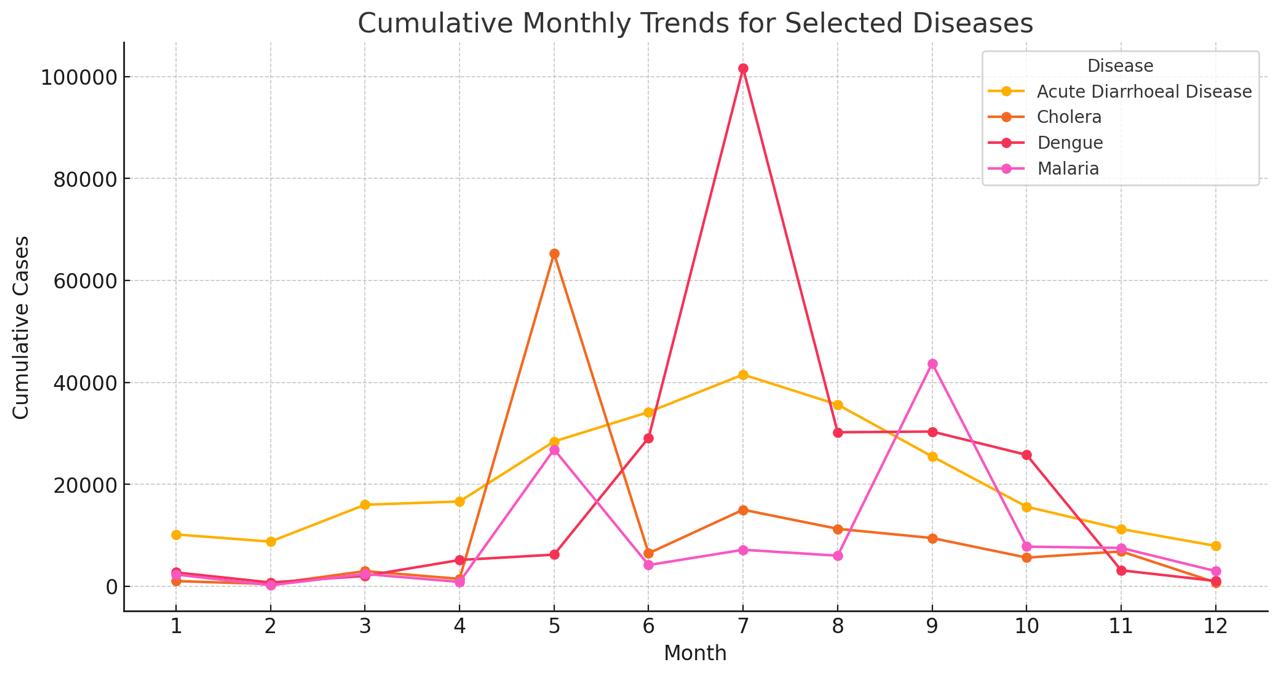}
    \caption{Cumulative monthly trend for selective diseases}
    \label{fig: cumulative monthly trend}
\end{figure}

\paragraph{}
Figure \ref{fig: cumulative monthly trend} The line graph illustrates cumulative monthly trends for selected diseases, including Acute Diarrheal Disease, Cholera, Dengue, and Malaria. The x-axis represents the months of the year (1-12), while the y-axis denotes the cumulative number of cases. The trends indicate seasonal variations, with significant spikes observed for dengue and cholera cases around mid-year, particularly in months 5-7. The dataset provides insights into how disease prevalence fluctuates over time, highlighting the impact of climatic factors on epidemic patterns.

\begin{figure}[h]
    \centering
    \includegraphics[width=0.5\textwidth]{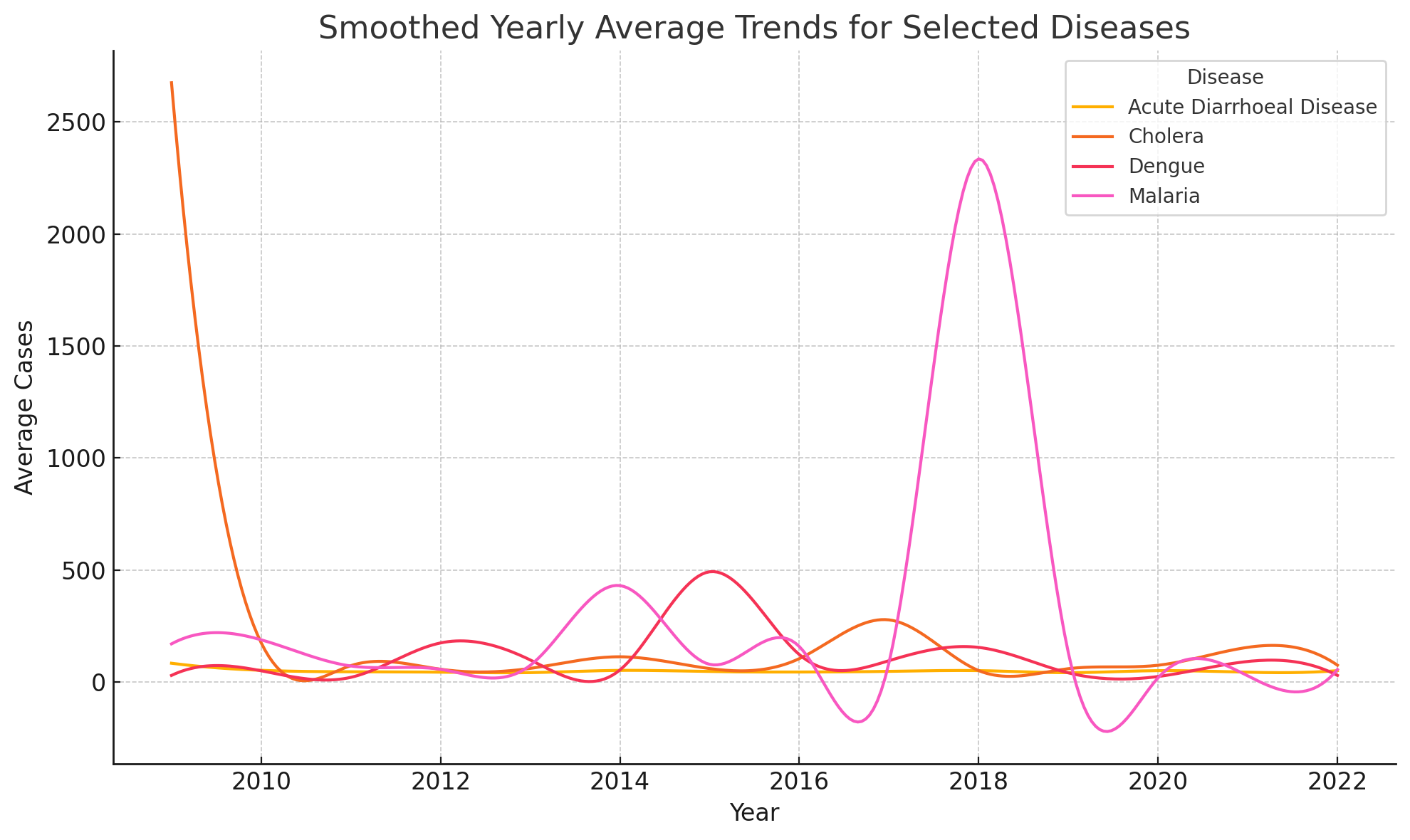}
    \caption{Smoothed yearly average trend for selective diseases}
    \label{fig: smoothed yearly average trend}
\end{figure}

\paragraph{Yearly Trends:}
Figure \ref{fig: smoothed yearly average trend} The smoothed yearly average trends in this Figure indicate long-term epidemiological variations, 
likely influenced by climate factors such as temperature, precipitation, and humidity. Dengue and malaria, both vector-borne diseases, exhibit periodic spikes, which could be linked to monsoon 
seasons that create favorable breeding conditions for mosquitoes. Cholera, a waterborne disease, 
saw a sharp decline post-2009 but still exhibits fluctuations, likely correlating with heavy rainfall 
events and water contamination patterns. Acute diarrheal disease appears relatively stable but 
may reflect underlying seasonal variations.

\begin{figure}[h]
    \centering
    \includegraphics[width=0.5\textwidth]{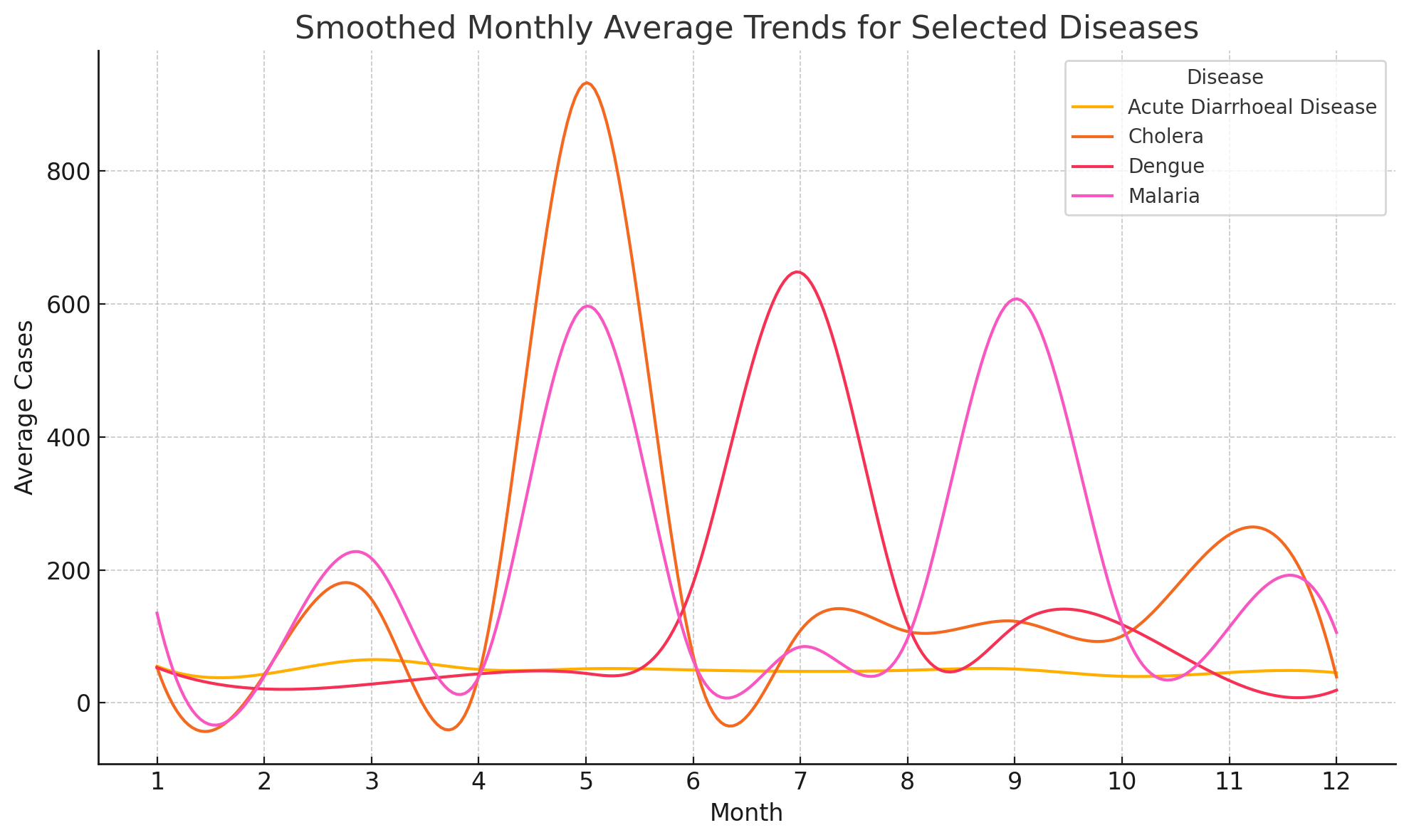}
    \caption{Smoothed months average trend for selective diseases}
    \label{fig: smoothed months average trend}
\end{figure}

\paragraph{Monthly Trends:}
 Figure \ref{fig: smoothed months average trend} The smoothed monthly average trends in this Figure provide insight into the seasonality of these diseases. Dengue and malaria cases surge around mid-year, particularly between May and September, aligning with monsoon rains, which promote mosquito breeding. Cholera exhibits peaks around similar periods, likely due to waterborne transmission during heavy rainfall events leading to water contamination. Acute diarrheal disease follows a more subdued but persistent trend throughout the year, with slight increases corresponding to seasonal weather changes that 
impact water and food safety.These trends highlight the strong climatological causation behind disease prevalence, demonstrating how environmental factors such as temperature and precipitation play a crucial role in shaping disease dynamics. Understanding these patterns is essential for integrating climate forecasts into epidemic prediction models, and enhancing preparedness and response 
strategies.

\begin{figure}[h]
    \centering
    \includegraphics[width=0.5\textwidth]{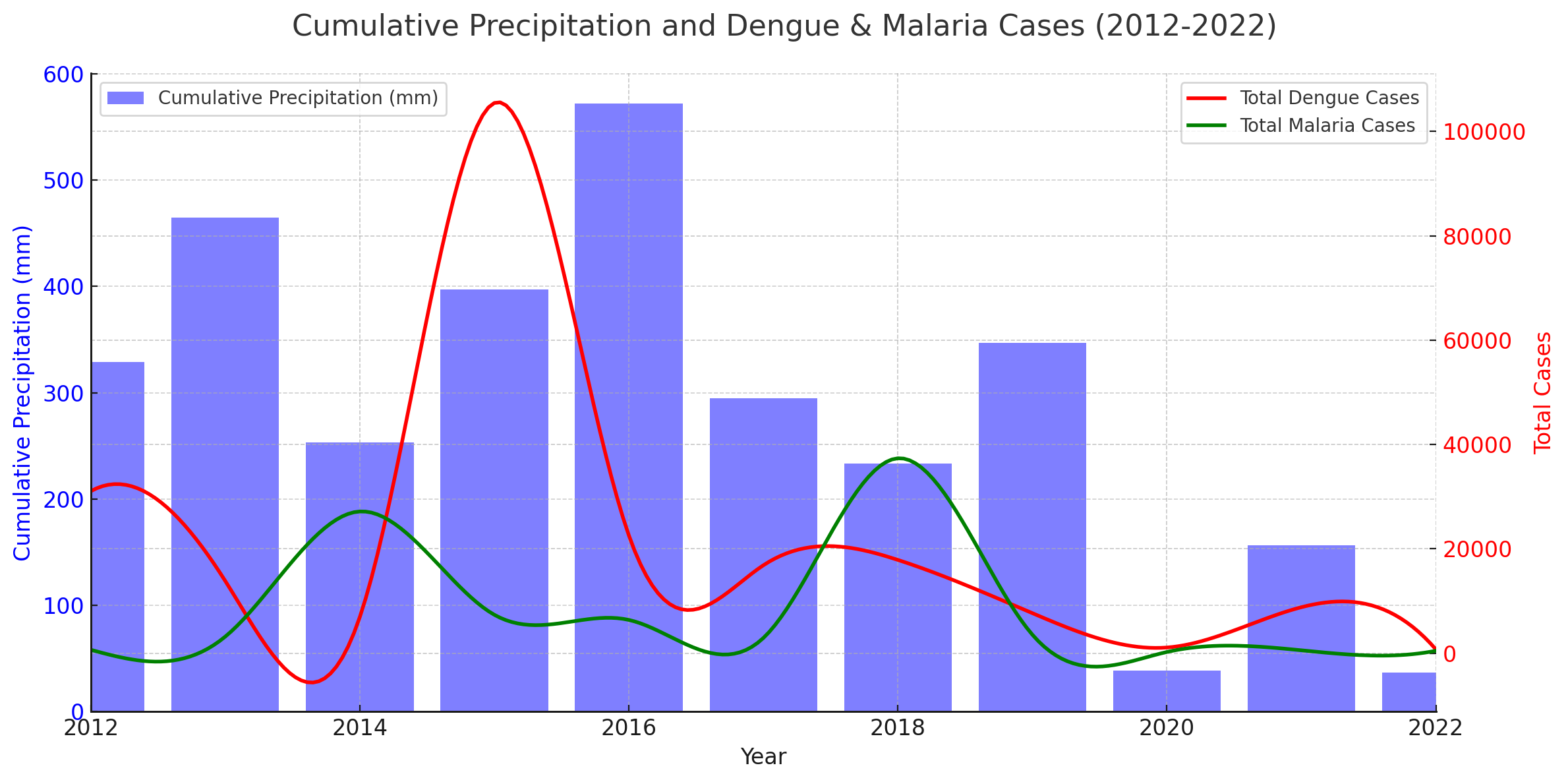}
    \caption{Cumulative Precipitation for Dengue and Malaria Cases(2012-2022)}
    \label{fig:ppt average}
\end{figure}

\paragraph{}
 The graph in Figures \ref{fig:ppt average} shows the connection between rainfall and the number of dengue and 
malaria cases over the years from 2012 to 2022. The blue bars represent cumulative precipitation, 
while the red and green lines show the total number of dengue and malaria cases, respectively. It 
is clear from the trends that years with high rainfall tend to see a rise in dengue cases, especially 
around 2016-2018 when dengue cases peaked significantly. This pattern makes sense because 
heavy rainfall creates pools of stagnant water, which provide ideal breeding grounds for Aedes 
mosquitoes, the main carriers of dengue. On the other hand, malaria cases follow a slightly 
different pattern, with fluctuations that are not as sharply linked to rainfall, possibly due to other 
factors like mosquito control efforts or environmental conditions that affect Anopheles 
mosquitoes, which spread malaria.
\paragraph{}
 The overall trend in the graph highlights how weather patterns, particularly rainfall, influence 
disease outbreaks. When there is more rain, water accumulates in different places, making it 
easier for mosquitoes to breed and spread diseases(Avramov, M., Thaivalappil, A., Ludwig, A., Miner, L., Cullingham, C. I., Waddell, L., \& Lapen, D. R. (2023).). This explains why outbreaks often follow wet 
years. However, the pattern is not always straightforward, as some years with high rainfall do not 
necessarily show a corresponding spike in cases. This could be due to improvements in public 
health measures, disease surveillance, or natural variations in mosquito populations. The graph 
serves as an important reminder of how climate and health are connected, emphasizing the need 
to use weather predictions in planning disease control strategies to prevent future outbreaks.

\subsection{Technical Validation}
We carried out statistical checks for completeness, consistency, and range integrity to verify the dataset, at all fields. The climatic variables in temperature, precipitation, and LAI are cross-checked for their accuracy from publicly available datasets. Disease case counts are verified with aggregated district and state-level health reports. The dataset is good for climate-health modeling from the granular level, its time span, and reliability. It further shows seasonal and regional variations of the disease that are highly associated with climatic factors. Therefore, the obtained correlations between temperature, precipitation, and epidemic cases indicate great possibilities for predictive modeling. This dataset is a fundamental basis for coupling climate forecasting with epidemic predictive models, and it opens a way to conduct proactive public health interventions.

\section{Results and Discussion}

\subsection{Temporal Trends of Disease Cases and Deaths}
\begin{figure}[h]
    \centering
    \includegraphics[width=0.7\columnwidth]{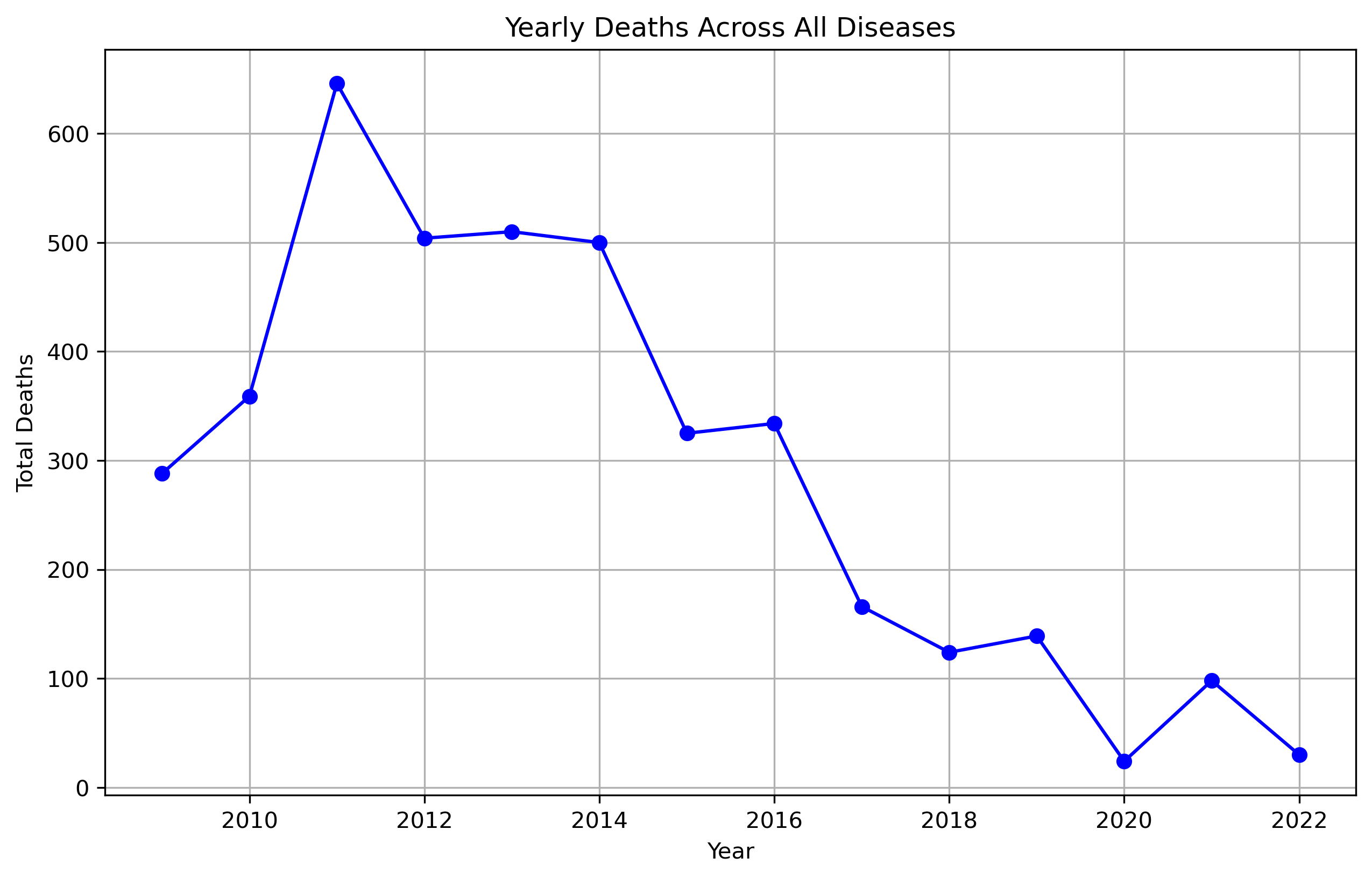}
    \caption{Yearly deaths across all diseases show a decreasing trend from 2010 to 2022.}
    \label{fig:yearly_deaths}
\end{figure}

Figure~\ref{fig:yearly_deaths} illustrates the temporal trends in yearly deaths across all diseases. A significant decline is observed from 2010 to 2022, suggesting improved public health interventions and increased awareness. Despite this decline, periodic surges in disease cases highlight the persistent seasonal nature of epidemics, necessitating continued monitoring and timely response.

\subsection{Spatial Distribution of Epidemic Diseases}
\begin{figure*}[h]
    \centering\includegraphics[width =0.6\textwidth]{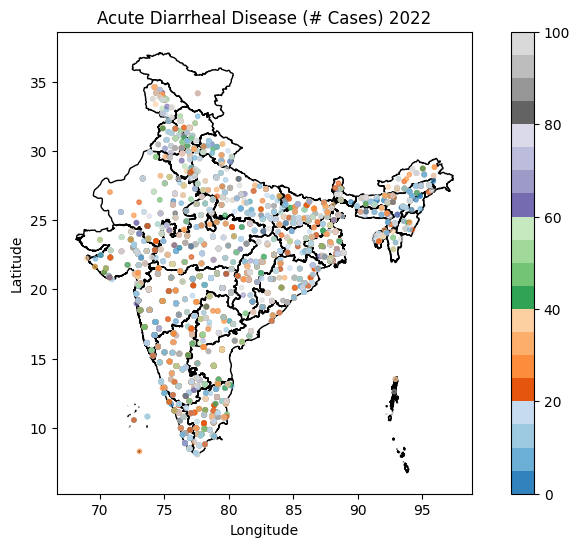}
    \caption{Spatial distribution of Acute Diarrhoeal Disease cases in India (2022)}
    \label{fig:add_2022}
\end{figure*}

The spatial distribution of Acute Diarrheal Disease cases in India (Figure~\ref{fig:add_2022}) highlights a high concentration of cases in northern and eastern states. This is consistent with regions that experience heavy monsoons and poor sanitation infrastructure, reinforcing the link between waterborne diseases and environmental conditions.
\clearpage

\subsection{Disease-Specific Trends}
\begin{figure*}[h]
    \centering
    \includegraphics[width=0.9\textwidth]{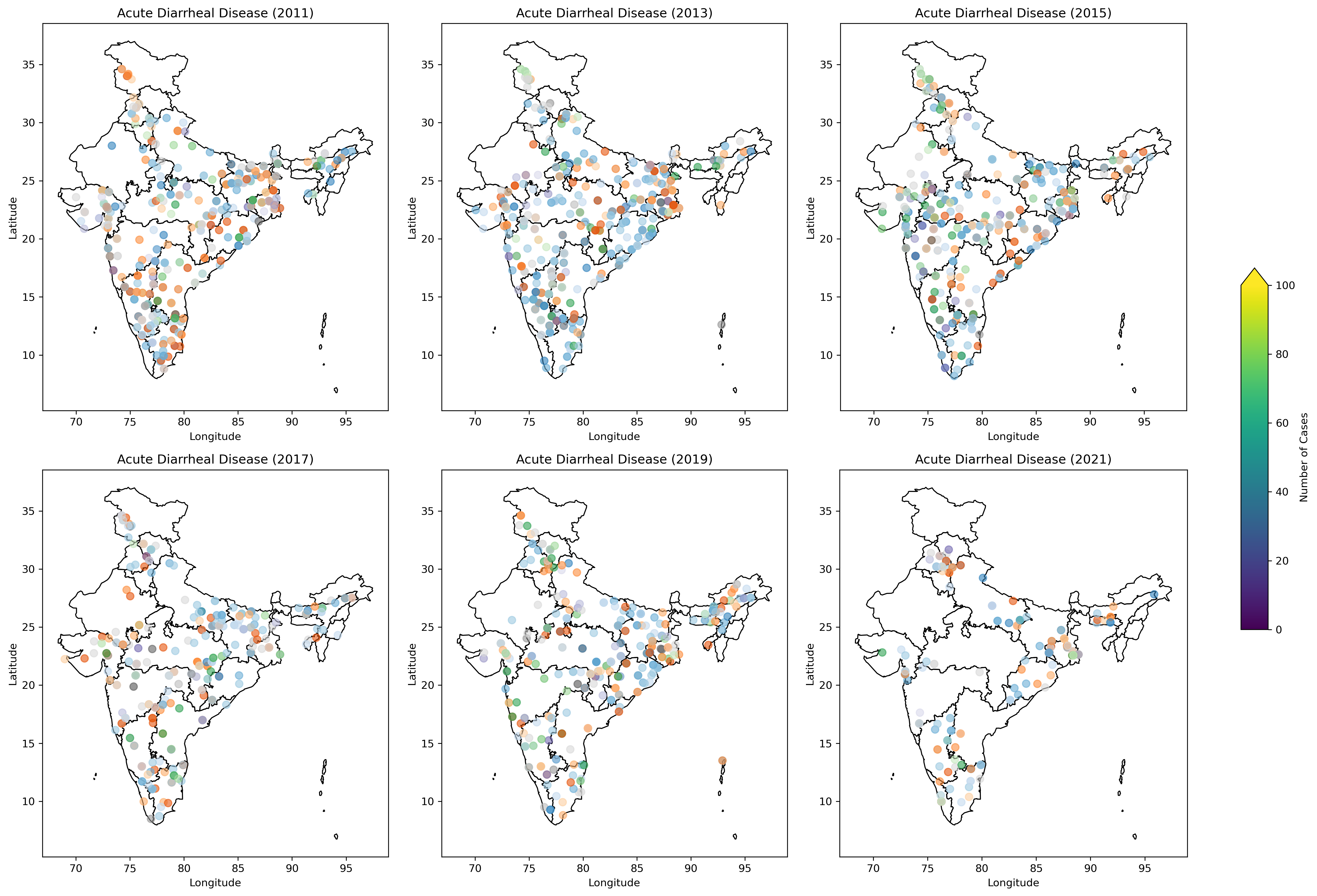}
    \caption{Temporal trends in Acute Diarrheal Disease cases across multiple years.}
    \label{fig:add_cases}
\end{figure*}

\begin{figure*}[h]
    \centering
    \includegraphics[width=0.9\textwidth]{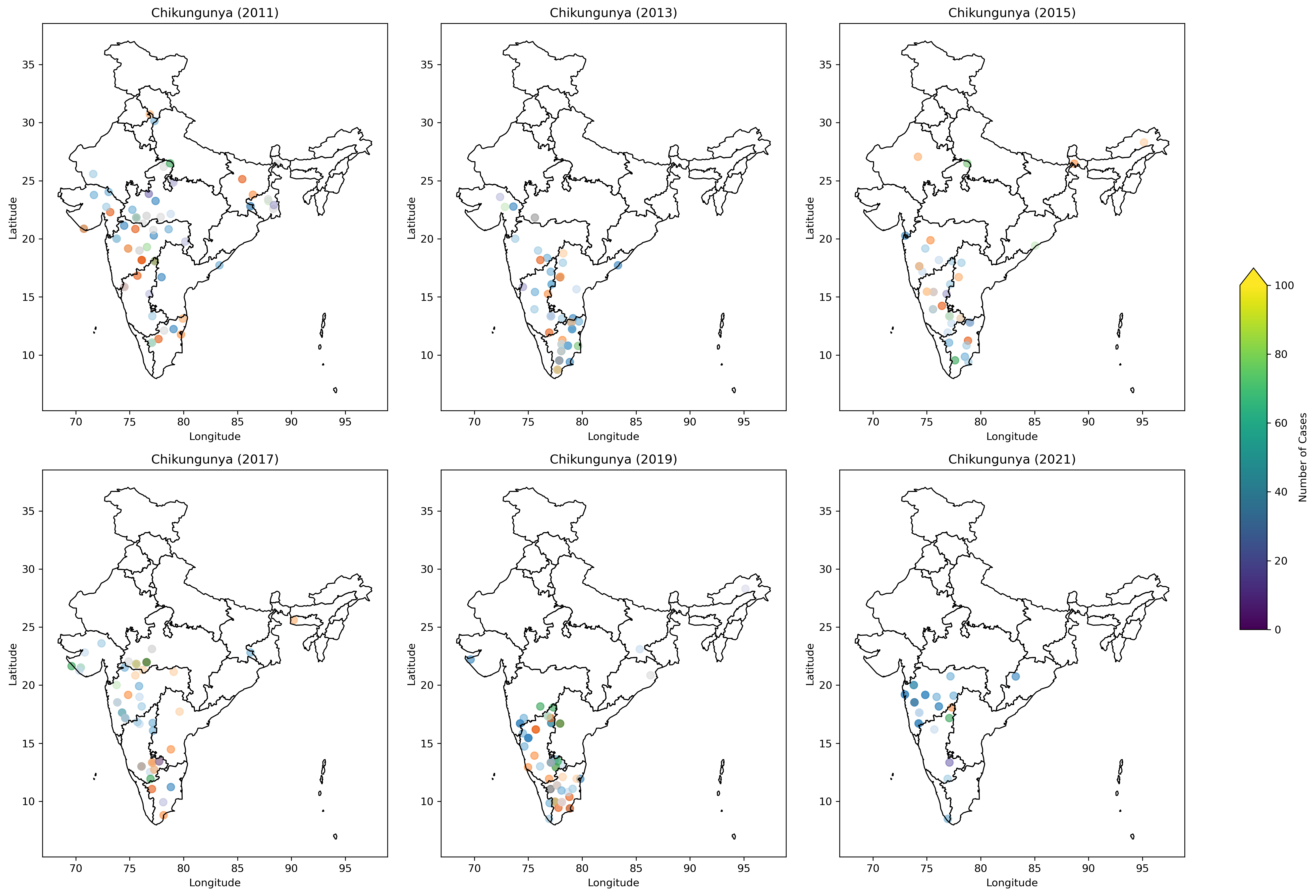}
    \caption{Spatial distribution of Chikungunya cases in India across selected years.}
    \label{fig:chikungunya_cases}
\end{figure*}

\begin{figure*}[h]
    \centering
    \includegraphics[width=0.9\textwidth]{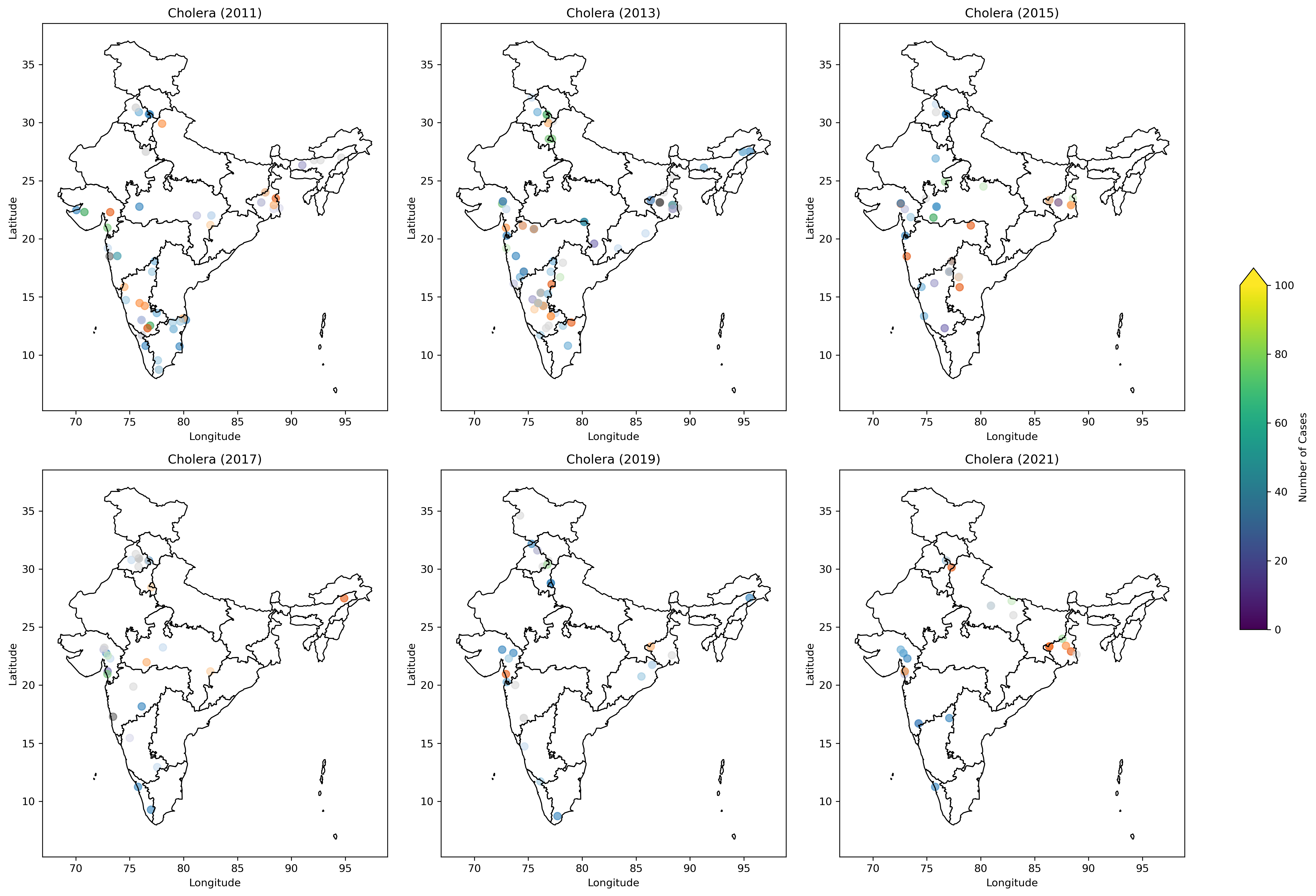}
    \caption{Spatial distribution of Cholera cases in India across selected years.}
    \label{fig:cholera_cases}
\end{figure*}

\begin{figure*}[h]
    \centering
    \includegraphics[width=0.9\textwidth]{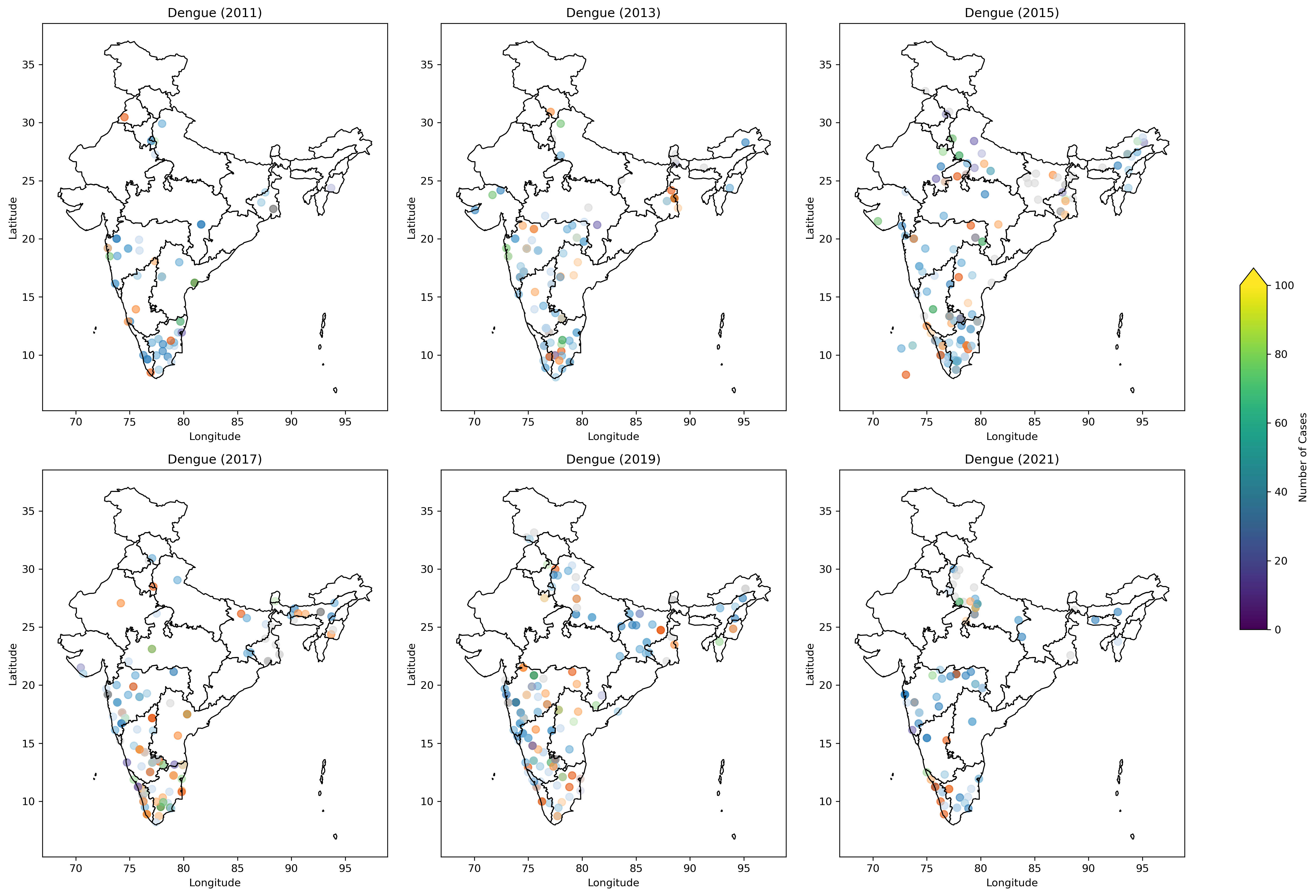}
    \caption{Spatial distribution of Dengue cases in India across selected years.}
    \label{fig:dengue_cases}
\end{figure*}

The temporal and spatial patterns of specific diseases are depicted in
Figures~\ref{fig:add_cases}, \ref{fig:chikungunya_cases}, \ref{fig:cholera_cases}, and \ref{fig:dengue_cases} . For Acute Diarrheal Disease (ADD), Figure~\ref{fig:add_cases}, shows a gradual increase in reported cases, peaking in 2019, with cases decreasing in the most recent years. This pattern can indicate the effectiveness of public health interventions, but some high-risk areas are still having hotspots. For Chikungunya, the pattern in Figure~\ref{fig:chikungunya_cases} shows irregular outbreaks dispersed in the southern and western regions as conditions for breeding remain favorable for the mosquitoes.
Trends of cholera cases, shown in Figure~\ref{fig:cholera_cases}, show a periodic outbreak mostly in coastal states which are known to experience heavy rainfall and flooding. Lastly, Fig. 8 shows the extensive geographical spread of Figure~\ref{fig:dengue_cases}, where the disease is constantly reported in several states. This points to the growing habitat range of Aedes mosquitoes, which are the disease's vectors.

\clearpage
\subsection{State-Wise Analysis}
\begin{figure}[h]
    \centering
    \includegraphics[width=0.8\columnwidth]{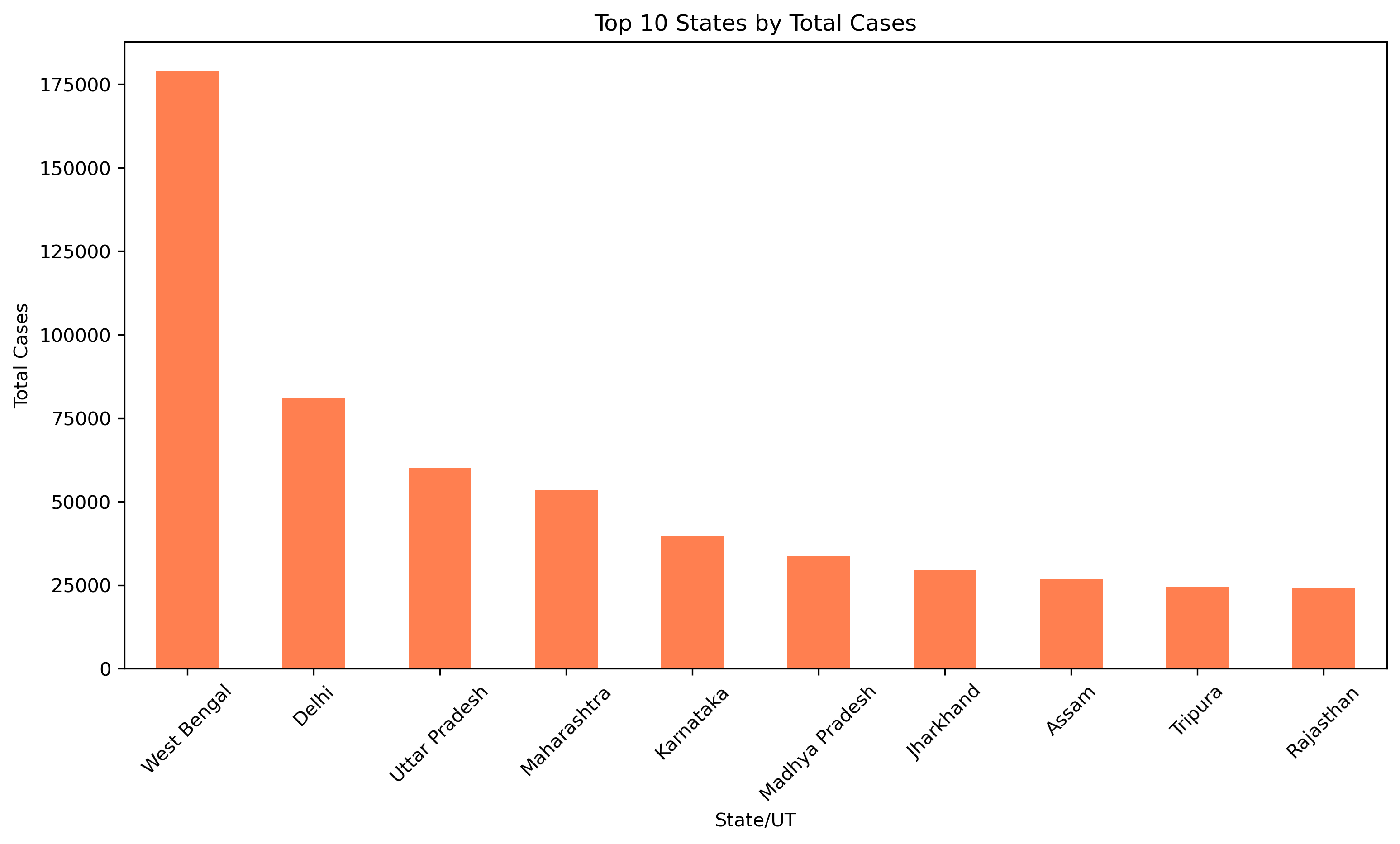}
    \caption{Indian States with the Highest Reported Disease Cases}
    \label{fig:top_states}
\end{figure}

 Figure~\ref{fig:top_states} ranks the top 10 states by total disease cases. West Bengal, Delhi, and Uttar Pradesh reported the highest disease burden, consistent with their high population densities and varying climatic conditions. Implementing targeted interventions that incorporate climate driven actions in these states can potentially lead to a substantial reduction in the national disease burden thereby significantly improving public health outcomes(Patz, J. A., \& Olson, S. H. (2006)).  

\section{Discussions \& Implications}

The findings underscore how climatic variables intersect with the incidence of diseases. Spatially and especially temporally, differ distinct patterns that are influenced by monsoons and temperature fluctuations and other factors such as urbanization. Acute Diarrhea Disease and Cholera have a huge impact from waterborne pathways; Dengue and Chikungunya have a strong association with the presence of mosquitoes.

Deaths have been trending downward, which means healthcare delivery has been improving; however, continuous outbreaks indicate prevention and early detection are still being overlooked. This dataset and related visualizations are actionables for policymakers to create interventions that can target specific improvements in sanitation and the implementation of climate-resilient public health strategies. Future work can involve machine learning models that can predict outbreaks from climatic variables to allow proactive mitigation and resource allocation.

\subsection{Spatial Hotspots}
Regions with high precipitation levels and dense vegetation, as indicated by the Leaf Area Index (LAI), exhibit a marked increase in the prevalence of vector-borne diseases such as malaria and dengue. These areas act as breeding grounds for disease vectors, underscoring the role of geographic and environmental factors in disease propagation.

\subsection{Temporal Patterns}
The data highlights seasonal surges in disease outbreaks, particularly during the monsoon months. This trend aligns with the proliferation of favorable conditions for vector growth, such as increased humidity and standing water. Such patterns demonstrate the seasonality of disease dynamics and the influence of short-term climatic changes.

\subsection{Climate-Disease Linkage}
Correlation analyses reveal a strong relationship between climatic variables, particularly temperature and precipitation, and the frequency of disease outbreaks. Higher temperatures, coupled with elevated moisture levels, contribute to vector activity and the subsequent rise in disease incidence.

These findings emphasize the critical need for integrating climate and health data to enhance the prediction and management of epidemic outbreaks. The spatial and temporal visualizations derived from this dataset offer invaluable insights for proactive public health planning, including targeted health interventions, resource allocation, and the development of robust climate-health models. By leveraging these insights, policymakers and researchers can better address the challenges posed by climate-induced epidemics, fostering resilience in public health systems.

\begin{figure}[h]
    \centering
    \includegraphics[width=0.8\textwidth]{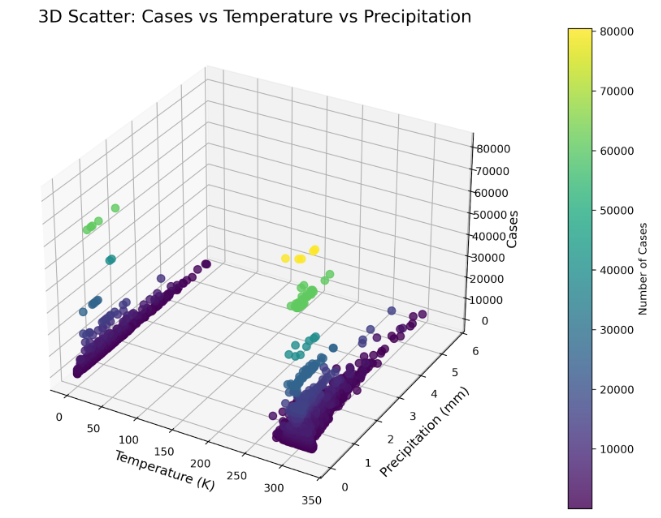}
    \caption{3D Scatter Plot of Disease Cases vs. Temperature and Precipitation. The color scale represents the number of cases, with warmer colors indicating higher case counts.}
    \label{fig:scatter3D}
\end{figure}

\vspace{0.5cm}
  Figure \ref{fig:scatter3D} reveals distinct clustering patterns of disease cases based on temperature and precipitation. Higher case counts are observed within specific ranges of temperature (approximately 300-350 K) and precipitation, suggesting the influence of these environmental factors on disease prevalence.  

\section{Results and Analysis}

This analysis points to a couple of time and environmental cycles that shape disease trends. The cumulative monthly trends indicate robust seasonality in four diseases: Acute Diarrhoeal Disease, Cholera, Dengue, and Malaria. Acute Diarrhoeal Disease and Cholera always correspond to high season, coinciding with the pre-monsoon and early monsoon, indicating a correlation to water contamination and poor sanitation. Conversely, spikes of Dengue and Malaria correlate to time frames when monsoon occurs and therefore coincides with mosquito-breeding convenience.

Smoothing the monthly averages smooths the patterns to give a clearer seasonal peaking for dengue, whereas malaria shows a broader post-monsoon rise. The two - waterborne and vector-borne diseases have marked differences of dependencies on climatic factors, thus bringing out environmental implications for the spread of both types of diseases.

Annual trends as seen through the cumulative precipitation and disease cases show considerable inter-annual variability.The peaks of 2014 and 2018 in the case of precipitation are observed with increased Malaria, so it is indeed direct. A sharp increase in Dengue cases in 2016 does not relate solely with the precipitation. Instead, changes in population or some vector control measure could be involved. The pattern followed by Dengue is an increasing trend till 2016 and a declining trend. This might have resulted from increased public health intervention measures. Malaria is trending downwards steadily, reflecting successful control efforts. Acute Diarrhoeal Disease and Cholera are trending more stably, indicating persistent issues related to water and sanitation.

In summary, the analysis highlights the fact that seasonality and precipitation play a very influential role in the occurrence of diseases. Some diseases like Malaria show progress in their control efforts, while others like Dengue call for vigilance and targeted interventions to minimize their impact. The findings are therefore crucial to integrate climatic and environmental monitoring into public health strategies.

\subsection{APPENDIX}
\begin{figure}[h]
    \centering
    \includegraphics[width=0.8\textwidth]{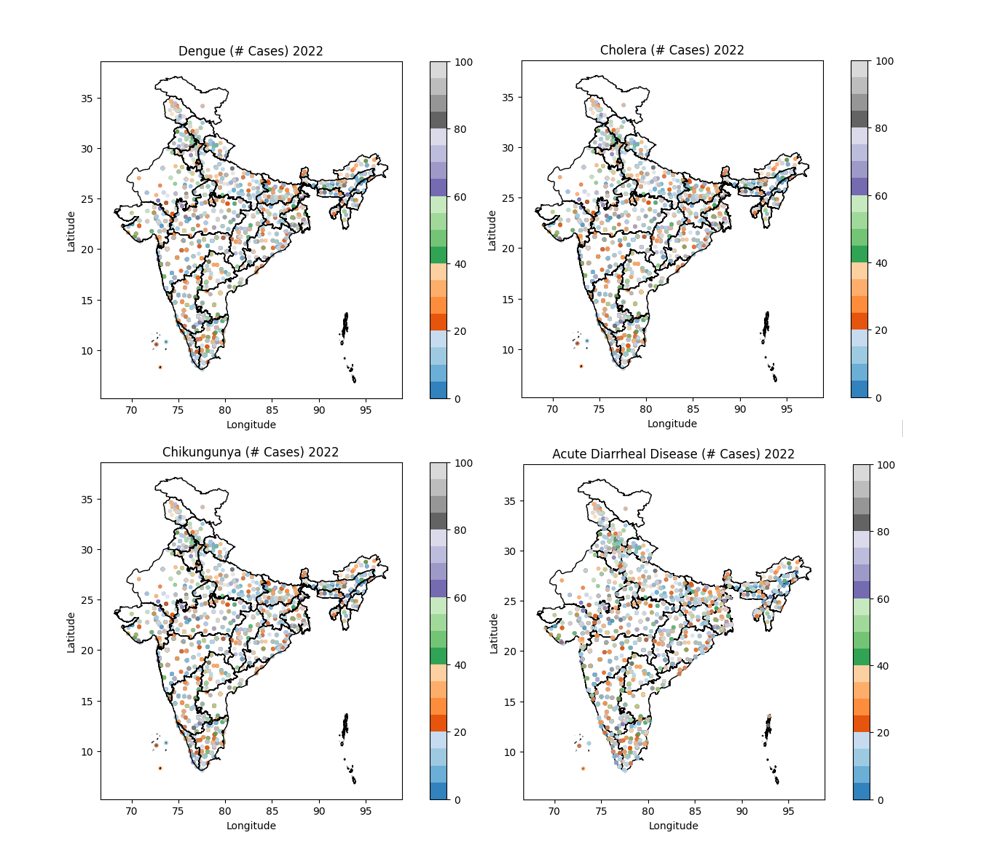}
    \caption{Disease incidence maps across India of four illnesses (2022)}
    \label{fig:appendix}
\end{figure}

The figure showcases the spatial distribution and prevalence of four key diseases—Acute Diarrheal Disease (ADD), Chikungunya, Dengue, and Cholera—across India for the year 2022. Each subplot represents one disease, with data visualized as colored dots indicating case counts at specific geographic locations. The legend provides a scale ranging from lower to higher cases, with darker shades denoting higher prevalence.

\subsubsection{Spatial Patterns and Disease-Specific Observations}

\paragraph{Acute Diarrheal Disease (ADD):}
The distribution of ADD cases is scattered throughout the country, with hotspots of high cases in the states with the densest population and the highest water quality problems. This distribution indicates that the sanitation and availability of water impact the prevalence of ADD in northern and eastern India. Concentration of cases in Uttar Pradesh and Bihar may indicate regional differences in public health infrastructure and the control of waterborne diseases.

\paragraph{Chikungunya:}
Chikungunya cases are less uniformly distributed compared to ADD, with higher concentrations observed in southern and western states, such as Karnataka and Maharashtra. This pattern is correlated with the climatic factors that favor the breeding of Aedes mosquitoes, such as warm temperatures and seasonal rainfall. The spatial variation underscores the need for region-specific vector control programs.

\paragraph{Dengue:}
Dengue has a much wider spread and thus has an equally high burden both in northern and southern regions. The fast-urbanizing cities, coupled with inadequate waste management, contribute to the mosquito-breeding friendly climate, hence Delhi and Mumbai, among other major cities, likely have very high cases. It is prevalent in different climatic zones, therefore the multifaceted nature, brought about by factors of urbanization and climatic factors.

\paragraph{Cholera:}
Most cases of cholera are located in the eastern and northeastern states like West Bengal and Assam. Those states are usually prone to floods and poor sanitation infrastructure. The clusters explain the convergence of climatic vulnerabilities, like monsoonal flooding, with public health challenges in those regions.

\subsubsection{Regional and Climatic Drivers}

The maps collectively emphasize the influence of climatic and geographic variability on disease prevalence. High-density population areas and regions prone to extreme weather events, such as flooding, emerge as hotspots for multiple diseases. Additionally, regions with warmer climates and seasonal rainfall patterns show a higher burden of vector-borne diseases like Dengue and Chikungunya.

\subsubsection{Implications for Public Health}

This spatial analysis highlights the critical need for targeted interventions and improved public health infrastructure. States with recurrent high disease burdens, such as West Bengal, Uttar Pradesh, and Delhi, should prioritize integrated disease management strategies, including improved sanitation, robust vector control measures, and early-warning systems tied to climate forecasts. Addressing regional vulnerabilities through climate-sensitive public health planning could significantly reduce the national disease burden and foster resilience against climate-induced health crises.

\newpage
\bibliographystyle{ieeetr} 

\bibliography{cas-refs}

\begin{enumerate}
    \item Patz, J. A., et al. (2005). "Impact of regional climate change on human health." Nature. \url{https://www.nature.com/articles/nature04188}
    \item Levy, K., Woster, A. P., Goldstein, R. S., \& Carlton, E. J. (2016). "Untangling the impacts of climate change on waterborne diseases: A systematic review of relationships between diarrheal diseases and climatic factors." Environmental Science \& Technology.
    \item Yang, K., et al. (2021). "Applications of AI in climate modeling and prediction." Patterns.
    \item \url{https://idsp.mohfw.gov.in/index.php} This web portal is for online reporting under Integrated Disease Surveillance Programme (IDSP), one of the major National Health Programme under the National Health Mission for all States \& UTs.
    \item Taylor, K. E., Stouffer, R. J., \& Meehl, G. A. (2012). "An overview of CMIP5 and the experiment design." Bulletin of the American Meteorological Society.
    \item India: Health of the Nation’s States – The India State-Level Disease Burden Initiative (2017). "Indian Council of Medical Research."
    \item Haines, A., et al. (2006). "Climate change and human health: Impacts, vulnerability, and public health." The Lancet.
    \item Bouma, M. J., \& Dye, C. (1997). "Cycles of malaria associated with El Niño in Venezuela." Journal of the American Medical Association (JAMA).
    \item Watts, N., et al. (2019). "The 2019 report of The Lancet Countdown on health and climate change: Ensuring that the health of a child born today is not defined by a changing climate." The Lancet, 394(10211), 1836-1878.
    \item Myers, S. S., et al. (2017). "Climate change and global food systems: Potential impacts on food security and undernutrition." Annual Review of Public Health, 38, 259-277.
    \item Rocklöv, J., et al. (2016). "Climate change and the rising burden of vector-borne diseases in Europe." Eurosurveillance, 21(16), 30204.
    \item Levy, K., et al. (2016). "Untangling the impacts of climate change on waterborne diseases: A systematic review of relationships between diarrheal diseases and temperature, rainfall, flooding, and drought." Environmental Science \& Technology, 50(10), 4905-4922.
    \item Patz, J. A., et al. (2014). "Climate change: Challenges and opportunities for global health." JAMA, 312(15), 1565-1580.
    \item World Health Organization (WHO). (2018). "Climate change and health." Retrieved from \url{https://www.who.int/news-room/fact-sheets/detail/climate-change-and-health}
    \item Ebi, K. L., et al. (2021). "Hot weather and heat extremes: Health risks." The Lancet, 398(10301), 698-708.
    \item Campbell-Lendrum, D., et al. (2015). "Climate change and health: Impacts, vulnerability, and adaptation." Annual Review of Public Health, 36, 293-312.
    \item Birkmann, J., et al. (2021). "Data science and climate change: Opportunities and challenges for integrated risk assessment." Nature Climate Change, 11(6), 485-494.
    \item Ostfeld, R. S. (2009). Climate change and the distribution and intensity of infectious diseases. Ecology, 90(4), 903-905.
\item Borghi, J., \& Kühn, M. (2024). A health economics perspective on behavioural responses to climate change across geographic, socio-economic and demographic strata. Environmental Research Letters. https://doi.org/10.1088/1748-9326/ad5d0c
    \item Nie, P., Zhao, K., Ma, D., Liu, H., Amin, S., \& Yasin, I. (2024). Global Climate Change, Mental Health, and Socio-Economic Stressors: Toward Sustainable Interventions across Regions. Sustainability, 16(19), 8693. https://doi.org/10.3390/su16198693
    \item Garcia, E., Eckel, S. P., Silva, S., McConnell, R., Johnston, J., Sanders, K. T., Habre, R., \& Baccarelli, A. (2024). The future of climate health research: An urgent call for equitable action- and solution-oriented science. Environmental Epidemiology, 8(5), e331. https://doi.org/10.1097/ee9.0000000000000331
    \item Athirsha, A., Anitha, M., Sundar, J. S., Kalpana, S., Valarmathi, S., \& Srinivas, G. (2024). Unleashing the relationship between climate change and infectious diseases. International Journal of Community Medicine and Public Health, 11(11), 4569–4576. https://doi.org/10.18203/2394-6040.ijcmph20243325
    \item Fairchild, G., Tasseff, B., Khalsa, H., Generous, N., Daughton, A. R., Velappan, N., Priedhorsky, R., \& Deshpande, A. (2018). Epidemiological Data Challenges: Planning for a More Robust Future Through Data Standards. Frontiers in Public Health, 6, 336. https://doi.org/10.3389/FPUBH.2018.00336
    \item Patz, J. A., \& Olson, S. H. (2006). Climate change and health: global to local influences on disease risk. Annals of Tropical Medicine and Parasitology, 100, 535–549. https://doi.org/10.1179/136485906X97426
    \item Avramov, M., Thaivalappil, A., Ludwig, A., Miner, L., Cullingham, C. I., Waddell, L., \& Lapen, D. R. (2023). Relationships between water quality and mosquito presence and abundance: a systematic review and meta-analysis. Journal of Medical Entomology. https://doi.org/10.1093/jme/tjad139

\end{enumerate}

\end{document}